# Magnetocaloric effect in $R_2Ti_3Ge_4$ (R = Gd, Tb and Er) Compounds

R. Nirmala

*Tata Institute of Fundamental Research, Mumbai 400 005, India*
*Department of Physics, Indian Institute of Technology Madras, Chennai 600 036, India*

Darshan C. Kundaliya

*Center for Superconductivity Research, Department of Physics, University of Maryland,*
*College Park, MD 20742, USA*
*Intematix Corporation, 46410 S. Fremont Blvd, Fremont, CA 94538*

A. V. Morozkin

*Department of Chemistry, Moscow Lomonosov State University, Moscow, 119899, Russia*

S. K. Malik[*]

*International Centre for Condensed Matter Physics - ICCMP, University of Brasilia, Brasilia, Brazil*

**Abstract**

Heat capacity of polycrystalline $R_2Ti_3Ge_4$ (R = Gd, Tb and Er) compounds (Orthorhombic, $Sm_5Ge_4$-type, Space group *Pnma*) has been studied in the temperature range of 1.8 K to 300 K in various applied magnetic fields. The compounds with magnetic lanthanide elements show interesting low field magnetism intrigued by possible presence of competing antiferromagnetic and ferromagnetic interactions. The magnetocaloric effect in these compounds is estimated from the field dependent heat capacity data. The magnetic entropy change and the adiabatic temperature change in the vicinity of the magnetic transition are found to be significant.

**Keywords:** Rare earth intermetallics and alloys, heat capacity, magnetocaloric effect.

*Corresponding author.

E-mail: skm@tifr.res.in

**Introduction**

Increasing demands to develop alternate and eco-friendly energy technologies have rejuvenated the syntheses and studies on magnetocaloric materials [1-3]. The intermetallic compound $Gd_5Si_2Ge_2$ and the related family of $R_5(Si_xGe_{1-x})_4$ (R = Rare earth)-type compounds are known to exhibit giant magnetocaloric effect (GMCE) over a wide temperature range making them useful for near-room temperature and low temperature magnetic refrigeration applications [4]. The key disadvantage of $Gd_5Si_2Ge_2$ is that this compound becomes brittle as it is thermally cycled through the coupled magnetic and structural transition which eventually leads to a substantial reduction in GMCE thus making it not suitable for practical applications. There have been several efforts to alloy this material to overcome this problem. In this context, we have partly substituted titanium at the rare earth site of the parent $R_5Ge_4$ compound. The $R_2Ti_3Ge_4$ (R = Gd, Tb, Dy, Ho, Er and Sc) compounds crystallize in the orthorhombic $Sm_5Ge_4$–type structure (Space group *Pnma*, No. 62) [5]. Earlier studies of magnetization on these compounds indicate that $Gd_2Ti_3Ge_4$ orders ferromagnetically at 32 K and the other compounds with magnetic rare earths order antiferromagnetically in the temperature range of 2 - 20 K [6]. The compound $Sc_2Ti_3Ge_4$ is non-magnetic down to 2 K. In the present work, we explore the possible applicability of these compounds in magnetic refrigeration at low temperatures, by studying their magnetocaloric properties by means of heat capacity measurements in various applied fields. We find that the $R_2Ti_3Ge_4$ compounds with R = Gd, Tb and Er exhibit considerable magnetocaloric effect with adiabatic temperature change ($\Delta T_{ad}$) of about 4 to 8 K for 5 T field change.

**Experimental details**

Polycrystalline samples of $R_2Ti_3Ge_4$ (R = Gd, Tb, Er and Sc) were prepared by arc melting under argon atmosphere starting from stoichiometric amounts of high purity elements. The samples were characterized by room temperature X-ray powder diffraction. Magnetization measurements were performed using a SQUID magnetometer [MPMS, Quantum Design, USA] in the temperature range of 1.8– 300 K in fields up to 5.5 T. For measurement of magnetization in zero-field-cooled (ZFC) state, the sample was cooled from the paramagnetic state in zero applied field and magnetization was measured while warming the sample. The magnetization data in the field-cooled (FC) state were collected while cooling the sample in a field. Heat capacity data were collected by employing the relaxation technique in the temperature range of 1.8 K- 300 K in applied fields up to 5 T or 9 T using a Physical Property Measurement System [PPMS Model 6000, Quantum Design, USA].

**Results and Discussion**

Room temperature X-ray powder diffraction experiments confirm the single phase nature of the samples. The compound $Gd_2Ti_3Ge_4$ orders ferromagnetically with a $T_C$ of ~32 K [6]. The $Tb_2Ti_3Ge_4$ compound shows antiferromagnetic-like cusp at ~18 K whereas $Er_2Ti_3Ge_4$ compound exhibits a tendency to order at about 1.8 K. The compounds with magnetic rare earths Tb, Dy and Ho were designated as antiferromagnets earlier owing to the cusp in their magnetization curves and because of the small negative values of paramagnetic Curie temperature. However, zero field cooled and field cooled magnetization data of these compounds show a substantial irreversibility below the magnetic ordering temperature [Figure not shown] typical of the presence of competing ferromagnetic and antiferromagnetic exchange interactions in compounds of such $Sm_5Ge_4$–type layered structure.

In order to understand the magnetocaloric effect in these compounds, heat capacity measurements were carried out in zero and applied magnetic field. Zero field heat capacity of $R_2Ti_3Ge_4$ (R = Gd, Tb and Er) compounds each reveal a peak close to the magnetic ordering temperature (labeled $T_P$, see Table 1), which eventually gets suppressed in applied fields of 5 T and 9 T [Fig. 1a-c]. The $Sc_2Ti_3Ge_4$ compound is non-magnetic down to 1.8 K. The isothermal magnetic entropy change ($\Delta S_m$) and the adiabatic temperature change ($\Delta T_{ad}$) for $R_2Ti_3Ge_4$ (R = Gd, Tb and Er) compounds have been computed from the heat capacity data, in the usual manner [7], utilizing the following thermodynamic relations:

$$\Delta S_m(T,H) = \int_0^T \frac{C(T',H) - C(T',0)}{T'} dT' \tag{1}$$

$$\Delta T_{ad}(T)_{\Delta H} \cong [T(S)H_f - T(S)H_i]_S \tag{2}$$

The variation of $\Delta S_m$ with temperature and the corresponding adiabatic temperature change for a field change of 5 T (and 9 T) are plotted in Figs. 2 and 3. A maximum value of $\Delta S_m$ of about ~ 5.7 J/mol K, ~4.6 J/mol K and 8.9 J/mol K has been obtained for 5 T field variation for the Gd, Tb and Er-based compounds, respectively. The corresponding peak values of $\Delta T_{ad}$ for a 5 T field change are ~ 4.4 K for $Gd_2Ti_3Ge_4$, ~ 3.8 K for $Tb_2Ti_3Ge_4$, and ~ 8.7 K for $Er_2Ti_3Ge_4$. Thus the values of $\Delta S_m$ and $\Delta T_{ad}$ are found to be reasonable. The magnetic entropy change is considerably large for Er-based compound probably because the magnetic transition occurs at very low temperature where it is not clouded by lattice entropy. Although a clear peak is not evidenced in magnetization data down to 1.8 K, prominent peak is observed at 2.9 K in the zero field heat capacity data of $Er_2Ti_3Ge_4$. The small glitch in heat capacity around ~50 K for the Gd- and Tb-based compounds could be from spurious 5:3 phase that is not detectable in X-ray diffraction data or could be intrinsic to the samples. Neutron diffraction experiments will be rewarding to understand the finer details of the magnetic structure of these compounds.

The low temperature shoulder and a small peak in the $\Delta S_m$ vs T curve of $Gd_2Ti_3Ge_4$ and $Tb_2Ti_3Ge_4$ compounds indicate possible field-induced effects on magnetization. This is in concordance with the bifurcation in ZFC – FC magnetization at low temperatures. Similar field-induced anomalies in magnetic entropy data have earlier been observed in isostructural $Dy_5Si_2Ge_2$ [8] and some Laves phase intermetallic compounds [9]. Recent neutron diffraction experiments on $Tb_2Ti_3Si_4$ compound suggest that this compound is a canted ferromagnet [10] with ferromagnetic (1 1 0) layers of Tb-moments in *ab* plane but with non-colinear moments in neighbour layers. Thus a remarkable similarity is seen between the Si-based $R_2Ti_3X_4$- (X = Si, Ge) compounds and the Si-rich $R_5(Si_xGe_{1-x})_4$ compounds since both of them are most-often ferromagnets while the crystal structure is a small variant of the other. However, we do not find any signature for a first order transition in either magnetization or heat capacity data suggesting a crystallographic transition in $R_2Ti_3Ge_4$ compounds whereas it is one of the essential ingredients that contributed to the giant magnetocaloric effect of $Gd_5Si_2Ge_2$.

**Conclusions**

Magnetic and heat capacity studies on $R_2Ti_3Ge_4$ (R = Gd, Tb and Er) compounds reveal interesting magnetism and associated moderate magnetocaloric effect. A maximum of about 4 to 8 K adiabatic temperature change has been obtained for these compounds for a field change of 5 T.

**Acknowledgement**

One of the authors, A.V.M. thanks the Russian Foundation for Basic Research for support in the form of a project No. N 06-08-00233-Á.

**Table I**

Magnetic ordering temperature ($T_N/T_C$), maximum magnetic entropy change ($\Delta S_m)_{max}$, maximum adiabatic temperature change ($\Delta T_{ad})_{max}$ for 5 T field change, peak in the zero field heat capacity data ($T_P$) for $R_2Ti_3Ge_4$ (R = Gd, Tb and Er) compounds.

| Compound | $T_N / T_C$ (K) | $(\Delta S_m)_{max}$ (J/mol K) | $(\Delta T_{ad})_{max}$ (K) | $T_P$ (K) |
|---|---|---|---|---|
| $Gd_2Ti_3Ge_4$ | 32 | 5.7 | 4.4 | 34.5 |
| $Tb_2Ti_3Ge_4$ | 18 | 4.6 | 3.8 | 17.5 |
| $Er_2Ti_3Ge_4$ | <2 | 8.9 | 8.7 | 2.9 |

**Figure Captions**

Fig. 1a-c.    Heat capacity vs. temperature for $R_2Ti_3Ge_4$ (R = Gd, Tb and Er) compounds in different applied fields.

Fig. 2a-c.    Magnetic entropy change, $\Delta S_m$, vs. temperature for $R_2Ti_3Ge_4$ (R = Gd, Tb and Er) compounds.

Fig. 3a-c.    Adiabatic temperature change, $\Delta T_{ad}$, vs. temperature for $R_2Ti_3Ge_4$ (R = Gd, Tb and Er) compounds.

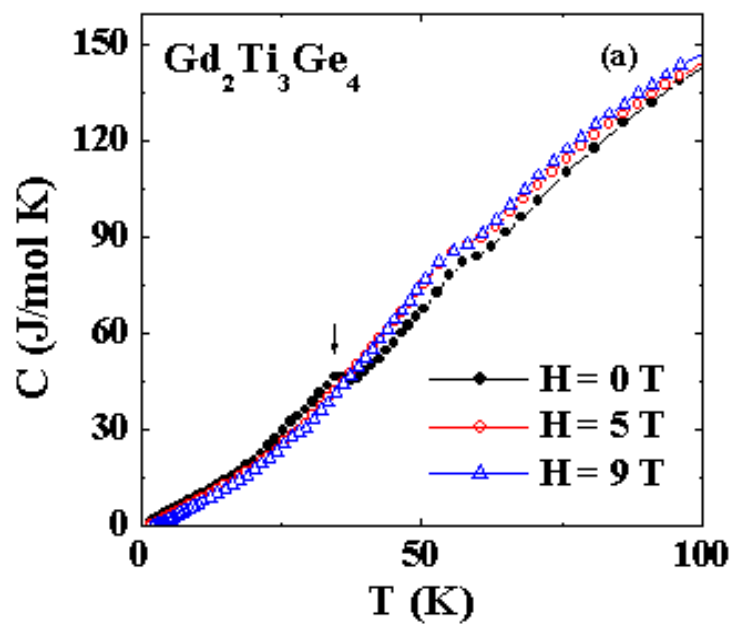

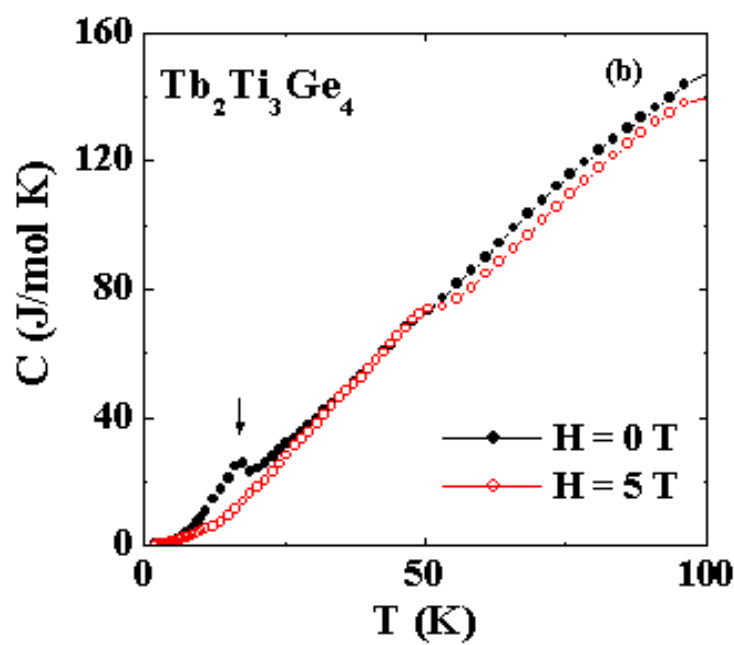

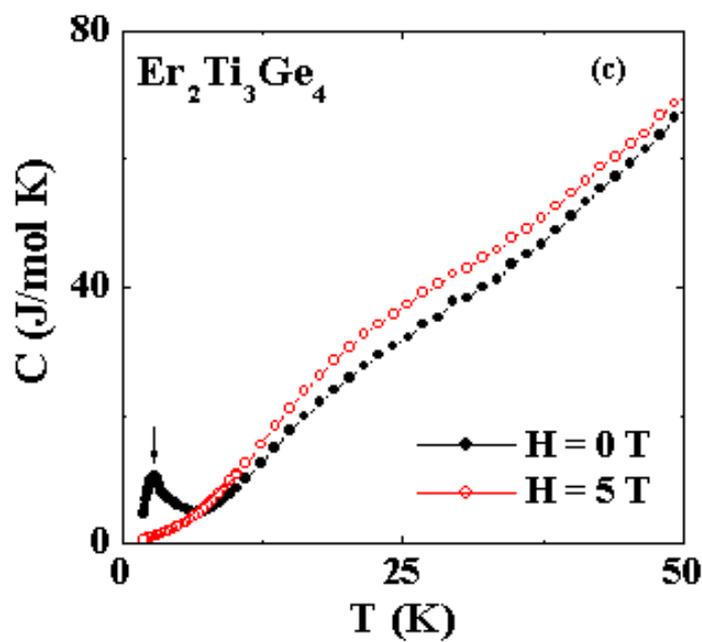

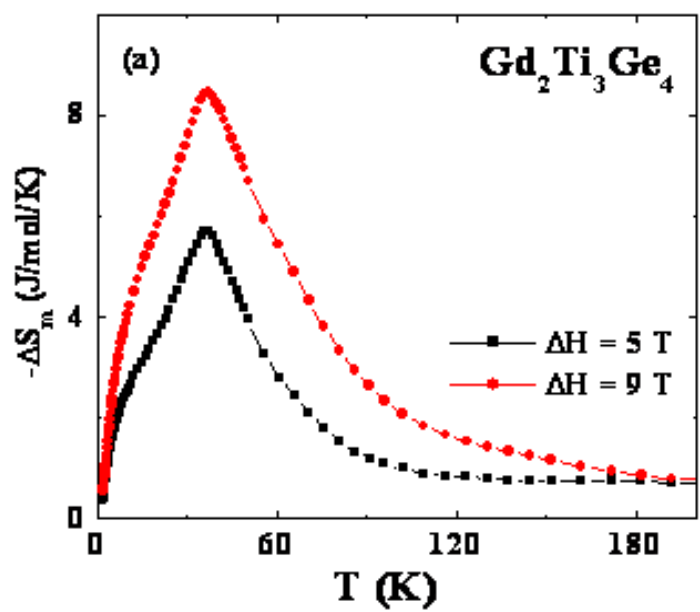

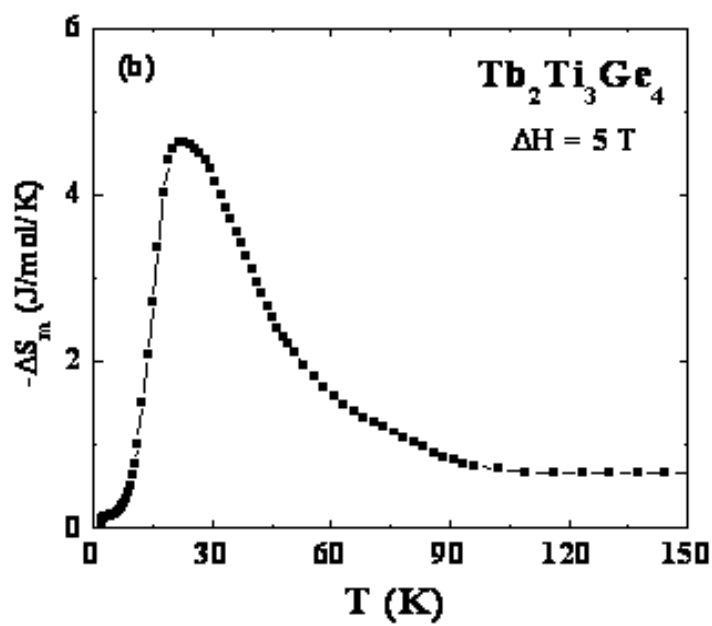

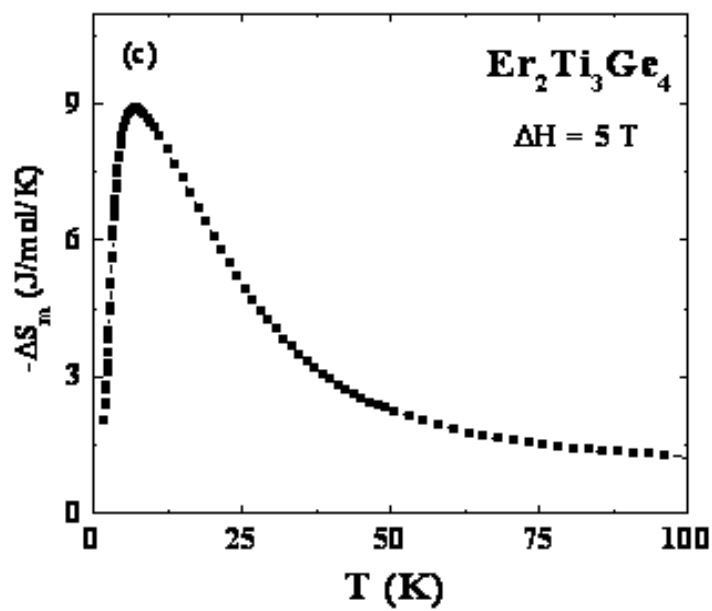

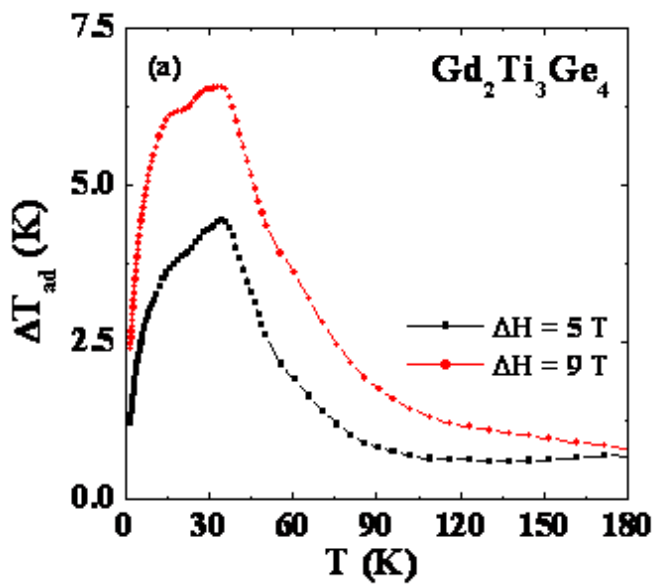
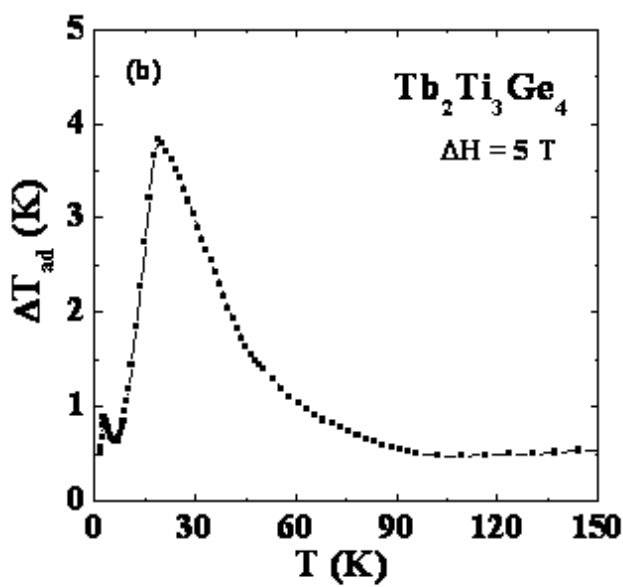
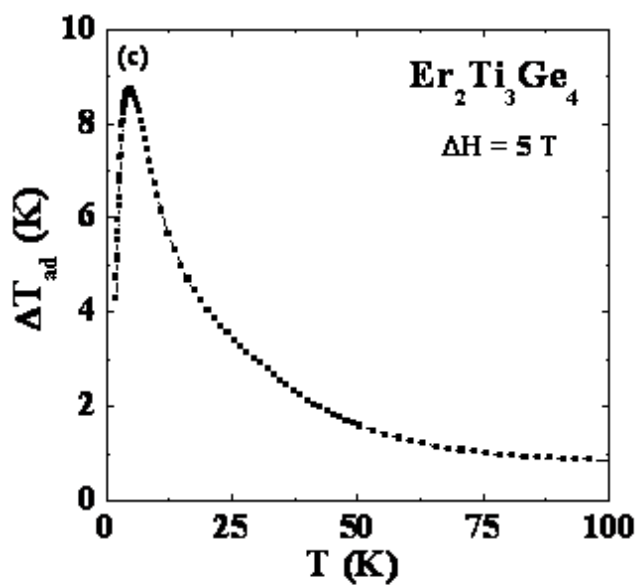